# Germanium monosulfide as a natural platform for highly anisotropic THz polaritons

Tobias Nörenberg[1,2], Gonzalo Álvarez-Pérez[3,4], Maximilian Obst[1], Lukas Wehmeier[1], Franz Hempel[1], J. Michael Klopf[5], Alexey Y. Nikitin[6,7], Susanne C. Kehr*[,1], Lukas M. Eng[1,2], Pablo Alonso-González*[3,4], and Thales V. A. G. de Oliveira*[1,2,5]

1 Institut für Angewandte Physik, Technische Universität Dresden, Dresden 01187, Germany

2 Würzburg-Dresden Cluster of Excellence - EXC 2147 (ct.qmat), Dresden 01062, Germany

3 Department of Physics, University of Oviedo, Oviedo 33006, Spain

4 Center of Research on Nanomaterials and Nanotechnology CINN (CSIC‑Universidad de Oviedo) El Entrego 33940, Spain

5 Institute of Radiation Physics, Helmholtz-Zentrum Dresden-Rossendorf, Dresden 01328, Germany

6 Donostia International Physics Center (DIPC), Donostia/San Sebastián 20018, Spain

7 IKERBASQUE, Basque Foundation for Science, Bilbao 48013, Spain

These authors contributed equally: Tobias Nörenberg and Gonzalo Álvarez-Pérez
Corresponding authors: *susanne.kehr@tu-dresden.de, *pabloalonso@uniovi.es, *thales.oliveira@tu-dresden.de

**Terahertz (THz) electromagnetic radiation is key to optically access collective excitations such as magnons (spins), plasmons (electrons), or phonons (atomic vibrations), thus bridging between optics and solid-state physics. Confinement of THz light to the nanometer length scale is desirable for local probing of such excitations in low dimensional systems, thereby inherently circumventing the large footprint and low spectral density of far-field THz optics. For that purpose, phonon polaritons (PhPs, i.e., light coupled to lattice vibrations in polar crystals) in anisotropic van der Waals (vdW) materials have recently emerged as a promising platform for THz nanooptics; yet the amount of explored, viable materials is still exiguous. Hence, there is a demand for the exploration of novel materials that feature not only THz PhPs at different spectral regimes, but also exhibit unique anisotropic (directional) electrical, thermoelectric, and vibronic properties. To that end, we introduce here the semiconducting alpha-germanium(II) sulfide (GeS) as an intriguing candidate. By employing THz nano-spectroscopy supported**

**by theoretical analysis, we provide a thorough characterization of the different in-plane hyperbolic and elliptical PhP modes in GeS. We find not only PhPs with long life times ($\tau > 2$ ps) and excellent THz light confinement ($\lambda_0/\lambda > 45$), but also an intrinsic, phonon-induced anomalous dispersion as well as signatures of naturally occurring PhP canalization within one single GeS slab.**

Polaritons are electromagnetic waves formed by light strongly coupled to collective excitations in matter.[1] The hybrid light-matter nature of polaritons offers a promising platform for the manipulation of the flow of light at the nanoscale.[2] Notably, phonon polaritons in layered vdW materials such as hBN, α-MoO$_3$ or α-V$_2$O$_5$ have recently attracted great interest[3–5] since, apart from featuring field confinement to the nanoscale, they naturally exhibit anisotropic (and particularly directional) propagation, ultra-long life times (of several ps), and low group velocities.[6] Polaritons hold great promises in manifold applications, such as for instance in nanolasers,[7,8] polarization-sensitive detectors,[9] molecular sensors,[10] quantum nano-photonics,[11,12] hyper-lensing,[13,14] waveguiding[15] and nano-optoelectronics.[16] However, a major challenge on the road to such applications is presented by the PhPs exclusively residing in the polar material's reststrahlen bands (RB): within the spectral region between the transverse optical (TO) and longitudinal optical (LO) phonon modes, typically located in the mid-infrared (MIR) to THz part of the electromagnetic spectrum, where the negative sign of the permittivity enables the excitation of polariton modes.[17,18] Thus, routes for spectral tunability (e.g. ion intercalation,[5] nano-structuring,[19] isotopic enrichment,[20] carrier photoinjection,[21] or dielectric environment[22,23]) as well as novel materials with RBs covering complementary spectral bands are of great need. Especially, in the scientifically and technologically emerging THz regime, the direct observation of PhP modes remains widely elusive with only few recent works.[24–27]

A promising material category to observe PhPs is provided by highly anisotropic vdW materials as they exhibit natural in-plane hyperbolic polariton dispersion resulting in ray-like propagation, enhanced confinement, and recently reported diffraction-less propagation in twisted-bilayer-engineered devices.[28–31] Furthermore, in contrast to in-plane hyperbolic PhPs in meta-materials, PhPs in natural vdW crystals exhibit significantly lower losses and are not limited by the precision of the fabrication.[6] Yet, the palette of vdW materials supporting nanoscale-confined PhPs in the THz spectral range is still very scarce. Here, we add a new member to this palette by introducing the family of group-IV monochalcogenides semiconductor compounds MX (M = Ge, Sn; X = S, Se) as a rich platform for THz nanophotonics. Their layered orthorhombic crystal structure - similar to black phosphorous[32] - gives rise to a large anisotropy of their optical, vibrational, and electrical properties.[33] All members of the material family show photovoltaic properties,[34] an enhanced thermoelectric effect,[34] large spin-orbit splitting,[35] high ionic dielectric screening,[36] and optical phonons well in the THz spectral range.[33] In this study, we focus on the recently predicted[37] THz PhPs in the compound alpha-germanium(II) sulfide (α-GeS, GeS) that exhibit an intriguing dispersion in the $\nu$ = 6.0 - 9.5 THz frequency range. Moreover, GeS has a direct bandgap of 1.6 eV, thus promising electric gating as a feasible approach for polaritonic control,[32,38] shows characteristic photoluminescence,[39] an outstanding

Seebeck coefficient possibly allowing for photocurrent-nanoscopy[40] of PhPs,[41] ferroelectricity in twisted nanowires[42,43] and in the monolayer limit,[44] resistance to oxidation[34] and fascinating exciton polaritons at visible wavelengths.[45]

The objective of the present work is the comprehensive characterization of the rich THz PhP modes including their dispersion, volume field distribution, quality factors, life times, and electromagnetic field confinement at the nanometer length scale. To that end, we carry out polariton interferometry experiments by employing a free-electron laser (FEL) as a narrowband THz light source. Our results, supported by full-wave simulations and transfer matrix and analytical calculations, unveil THz PhPs with large quality factors (Q = 10), long life times ($\tau$ > 2 ps), and deep subwavelength confinement (up to $\lambda_0/45$, where $\lambda_0$ is the incident free-space light wavelength), together with signatures of a natural PhP canalization regime.

The layered orthorhombic crystal structure of GeS (space group Pcmn) is depicted in **Figure 1**a. In analogy to that of black phosphorous,[32] it consists of covalently-bound layers stacked in [001]-direction with an armchair structure in the [100]-direction and zigzag structure in the [010]-direction.[33] The difference in lattice constants (a = 4.29 Å, b = 3.64 Å, c = 10.42 Å) is remarkable, in particular the large ratio a/b = 1.18 gives rise to a high structural anisotropy within the layers that is 2.5-times higher than in the likewise biaxial ($\varepsilon_x(\omega) \neq \varepsilon_y(\omega) \neq \varepsilon_z(\omega)$) $\alpha$-MoO$_3$ crystal (a/c = 1.072)[46] in which THz PhPs have been recently demonstrated.[24] Our micro-Raman spectrum (Figure 1b) unveils four Raman peaks at Raman shifts of 112, 213, 240, and 270 cm$^{-1}$ that can be readily attributed to the $A_g^3$, $B_{3g}$, $A_g^1$, and $A_g^2$ phonon modes, respectively.[39] Particularly, their polarization-dependence allows deducing the GeS crystal structure orientation of individual flakes. To that end, we find the maximum Raman intensity of the $A_g^3$ mode in Figure 1c (violet; that is parallel to the [100] crystal axis[47]) to be aligned along the right edge of the GeS flake marked in the optical microscopy image in Figure 1d.

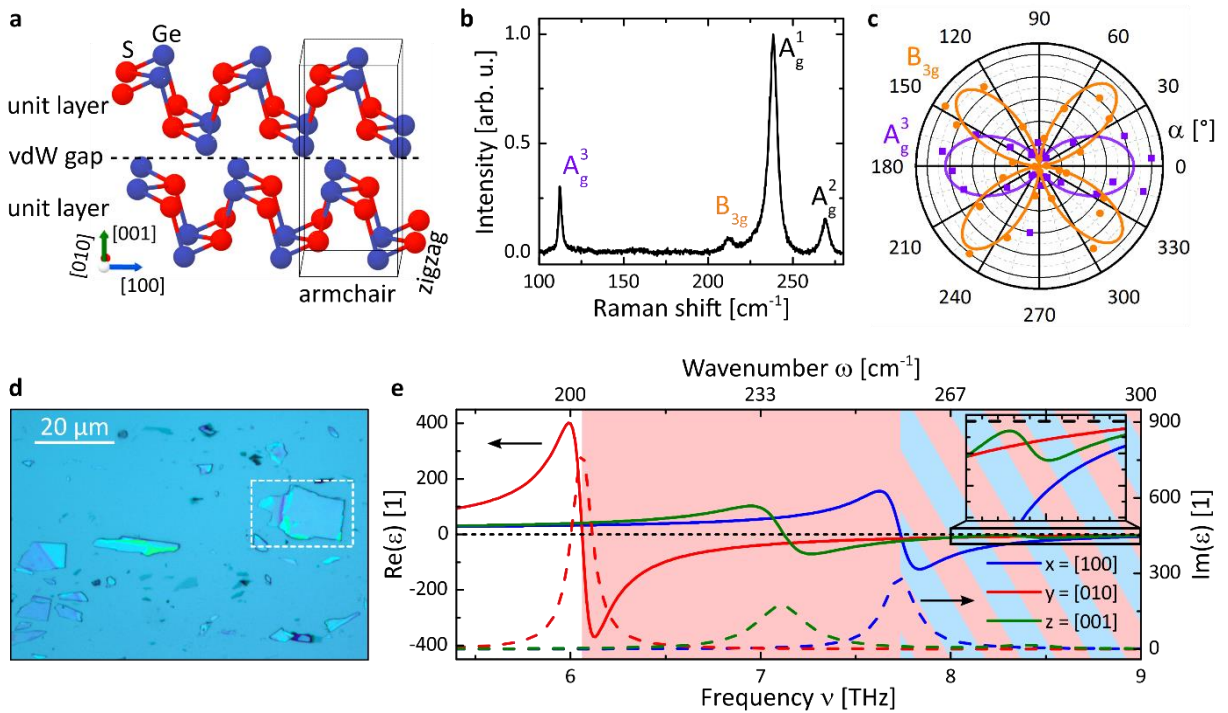

**Figure 1. Material properties of germanium sulfide ($\alpha$-GeS). a)** Crystal structure of $\alpha$-GeS. The crystal is composed of layers of covalently bounded Ge (blue) and S (red) atoms with the vdW stacking direction along the [001] crystal direction. Similar to black phosphorous the layers show an armchair and zigzag geometry along the [100] and [010] crystal directions, respectively. The box marks the unit cell. **b)** Representative Raman spectrum for an incident linear polarization rotated by 4° relative to the [100] crystal axis. The characteristic polarization dependence of the $A_g^3$ mode at 122 cm$^{-1}$ (violet) and $B_{3g}$ mode at 213 cm$^{-1}$ (orange) can be used for unambiguous identification of the crystal axis orientation. **c)** Polar plot of the normalized Raman scattering intensities of the $A_g^3$ and $B_{3g}$ mode in (b) with $\alpha = 0°$ corresponding to the [100] crystal direction. The lobes of the two-fold rotational symmetric $A_g^3$ mode extend along the [100] crystal direction.[47] The four-fold symmetric $B_{3g}$ mode is phase-shifted by 45° relative to the $A_g^3$ mode. **d)** Optical microscopy image of exfoliated GeS crystals on silicon. The dashed box marks the flake investigated in this work. **e)** Real (solid lines) and imaginary (dashed lines) parts of the complex permittivity $\varepsilon$. The permittivity in the THz regime is governed by four optical phonons and exhibits two (overlapping) in-plane reststrahlen bands $RB_y$ and $RB_x$ (shaded red and blue, respectively). The inset highlights the real part of $\varepsilon$ from 8 to 9 THz.

In addition to the Raman-active phonons, polar GeS exhibits several well-characterized, directional transversal optical (TO) phonons located in the THz spectral regime[33] that govern its dielectric permittivity $\varepsilon$ (Figure 1e). We define the coordinate system to align with the GeS crystallographic axes as x = [100], y = [010] and z = [001]. At frequencies from 6 to 10 THz, a negative permittivity (Re($\varepsilon_i$) < 0, i = x, y, z) is found along different crystal axes within four RBs with two of them defined in the plane: $RB_y$ ($\nu_{TO,y}$ = 6.06 and $\nu_{LO,y}$ = 9.47 THz) and $RB_x$ ($\nu_{TO,x}$ = 7.74 and $\nu_{LO,x}$ = 9.65 THz) shaded red and blue, respectively (note the large overlap as marked by the hatching). In z-direction, GeS exhibits two out-of-plane TO phonons ($\nu_{TO,z,1}$ = 7.1 THz and ($\nu_{TO,z,2}$ = 8.4 THz) that spectrally overlap with the in-plane TO phonons, giving rise to an exotic, highly anisotropic optical response.

To experimentally study the excitation of PhPs in GeS within these RBs, we performed s-SNOM-based polariton interferometry applying a narrowband, tunable FEL.[24] The experimental setup is sketched in **Figure 2**a: the pulsed THz radiation produced by the FEL (repetition rate 13 MHz, pulse duration > 5 ps) is focused on a metallized atomic force microscopy (AFM) tip that acts as a nanoantenna providing high k-vectors along with an enhanced, localized electric field. The polarized tip on top of the GeS flake launches PhPs that propagate away from the tip and are back reflected at edges of the 224 nm-thick flake. The electric field of the back-travelling PhPs is scattered by the same tip into the far-field, where it can be detected (see Supporting Information for details on the setup). By raster scanning the tip across the flake at a fixed incident frequency we get a spatial near-field (NF) $S_{2\Omega}$ image of the polaritons' interference pattern.[48] To ensure that our near-field images are recorded in an area with a homogeneous flake thickness and sharp edges, we restrict our s-SNOM measurements to the front-facing 90°-corner in Figure 2a (equivalent to the bottom right corner in Figure 1d). In order to corroborate the existence of propagating (Re[$k$] > Im[$k$]) PhPs in GeS, we carried out full-wave electromagnetic simulations at representative excitation frequencies within the in-plane reststrahlen bands: $RB_x$ and $RB_y$. More specifically, we simulate the electromagnetic fields generated by a vertical point dipole, in analogy to the AFM tip. The result for a frequency of ν = 7.33 THz within $RB_y$ is shown in Figure 2b (color plot), confirming an exotic polaritonic field distribution: starting from the exciting dipole located in the center of Figure 2b, a polariton propagates along the y (=[010])-direction with hyperbolic wave-fronts. Notably, no PhPs with wavevectors along the x (=[100])-direction are allowed, while the PhPs propagating along the y-direction have momenta $k_{y,sim}^{7.33} = (2.42 + 0.26i) \times 10^4$ cm$^{-1}$ ($\lambda_{y,sim}^{7.33} = 2.60$ μm). The latter are more than 15-times higher than the free-space momentum $k_0 \approx 1.54 \times 10^3$ cm$^{-1}$ ($\lambda_0 = 40.9$ μm), thus implying a considerable THz light confinement. The hyperbolic character of the excited PhPs is further supported by their analytically-calculated isofrequency curves (IFC, a section of the dispersion surface for a constant frequency) overlaid with the simulation in Figure 2b that exhibit the shape of open hyperbolas with its major axis aligned along the y-direction. Note that the opening angle of PhPs propagation in real space is well reproduced by the direction of the group velocity ($v_g$, dashed arrow in Figure 2b) in momentum space, which is oriented perpendicular to the IFCs (compare dashed line parallel to $v_g$). The IFCs were obtained using the recently derived equation for the polariton in-plane wavevector $k^2 = k_\parallel^2 = k_x^2 + k_y^2$ in a biaxial slab[15,49]

$$k(\omega) = \frac{\rho}{d}\left(\arctan\left(\frac{\varepsilon_1 \rho}{\varepsilon_z}\right) + \arctan\left(\frac{\varepsilon_3 \rho}{\varepsilon_z}\right) + \pi l\right), l = 0,1,2 \ldots \qquad (1)$$

with the slab thickness $d$, the permittivity of the superstrate (substrate) $\varepsilon_1$ ($\varepsilon_3$), the mode quantization index $l$, the GeS permittivity tensor diagonal elements $\varepsilon_x$, $\varepsilon_y$, $\varepsilon_z$, the angle between $k$ and the x-axis $\varphi$, and with $\rho = i\sqrt{\varepsilon_z/(\varepsilon_x \cos^2\varphi + \varepsilon_y \sin^2\varphi)}$. For a frequency of ν = 7.33 THz, this analytical model gives a polariton momentum in [010]-direction of $k_{y,cal}^{7.33} = (2.3 + 0.23i) \times 10^4$ cm$^{-1}$ ($\lambda_{y,cal}^{7.33} = 2.73$ μm) that excellently matches our numerical estimate.

The simulated field distribution Re($E_z$) for a second frequency ν = 8.57 THz that is located within the overlap between GeS in-plane reststrahlen bands $RB_x$ and $RB_y$, is displayed in Figure 2c. Here, the real parts of the corresponding permittivities Re($\varepsilon_x$) and Re($\varepsilon_y$) are both negative and, moreover, show a high degree of GeS crystal anisotropy Re($\varepsilon_x$)/Re($\varepsilon_y$) = 3.4. Consistently, the spatial distribution of Re($E_z$) reveals an elliptically propagating polariton with largely different wave-vectors along the in-plane crystal directions. The simulation predicts the [100]-crystal axis to be the prominent propagation direction with $k_{x,sim}^{8.57} = (3.31 + 0.93i) \times 10^4$ $cm^{-1}$ ($\lambda_{x,sim}^{8.57} = 1.90$ μm), accompanied by propagation along the [010]-direction with a about twofold higher momentum $k_{y,sim}^{8.57} = (6.49 + 4.1i) \times 10^4$ $cm^{-1}$ ($\lambda_{y,sim}^{8.57} = 0.97$ μm), albeit higher damping. These numerical findings are supported by the analytical IFCs in Figure 2c that are propeller-shaped with high (low) momenta along y (x)-direction. From **Equation (1)** we obtain the complex in-plane momenta $k_{x,cal}^{8.57} = (3.0 + 1.1i) \times 10^4$ $cm^{-1}$ and $k_{y,cal}^{8.57} = (8.2 + 7.0i) \times 10^4$ $cm^{-1}$ ($\lambda_{x,cal}^{8.57} = 2.09$ μm and $\lambda_{y,cal}^{8.57} = 0.77$ μm), respectively, which agree well with the numerical values. Intriguingly, the shape of the IFC and the field distribution in Figure 2c strongly imply a natural canalization along the [100]-direction at ν = 8.57 THz: the opening angle of propagation as observed in the color plot combined with the very low angular spread of the group velocities given by the IFC closely resemble the canalization of PhPs recently found in twisted bilayers of α-MoO$_3$.[28–31] A more detailed discussion of the natural canalization effect will be given below in the context of the dispersion as well as in the Supporting Information.

The optical near-field S$_{2\Omega}$ image recorded at an illuminating frequency ν = 7.33 THz (Figure 2d) shows polaritonic features matching the predicted in-plane hyperbolicity. The flake's near-field response containing the polariton's Re($E_z$) features characteristic fringes with a periodicity of half the polariton wavelength parallel to the horizontal flake edge that are caused by PhP propagation along [010]-direction.[4] In contrast, similar fringes parallel to the vertical edge are absent, thus confirming the in-plane hyperbolic character of the PhPs at this frequency. In large contrast, at ν = 8.57 THz the recorded near-field image in Figure 2e shows polaritonic fringes parallel to both flake edges. In particular, the fringe spacing (0.5 λ$_p$) parallel to the vertical edge is considerably larger than parallel to the horizontal edge. Moreover, while up to three distinct fringes are clearly visible decaying along the [100]-direction, the fringes decaying along the [010]-direction vanish quickly with distance from the flake edge with only a few fringes remaining barely visible. These observations are consistent with the corresponding complex momenta estimated from theory.

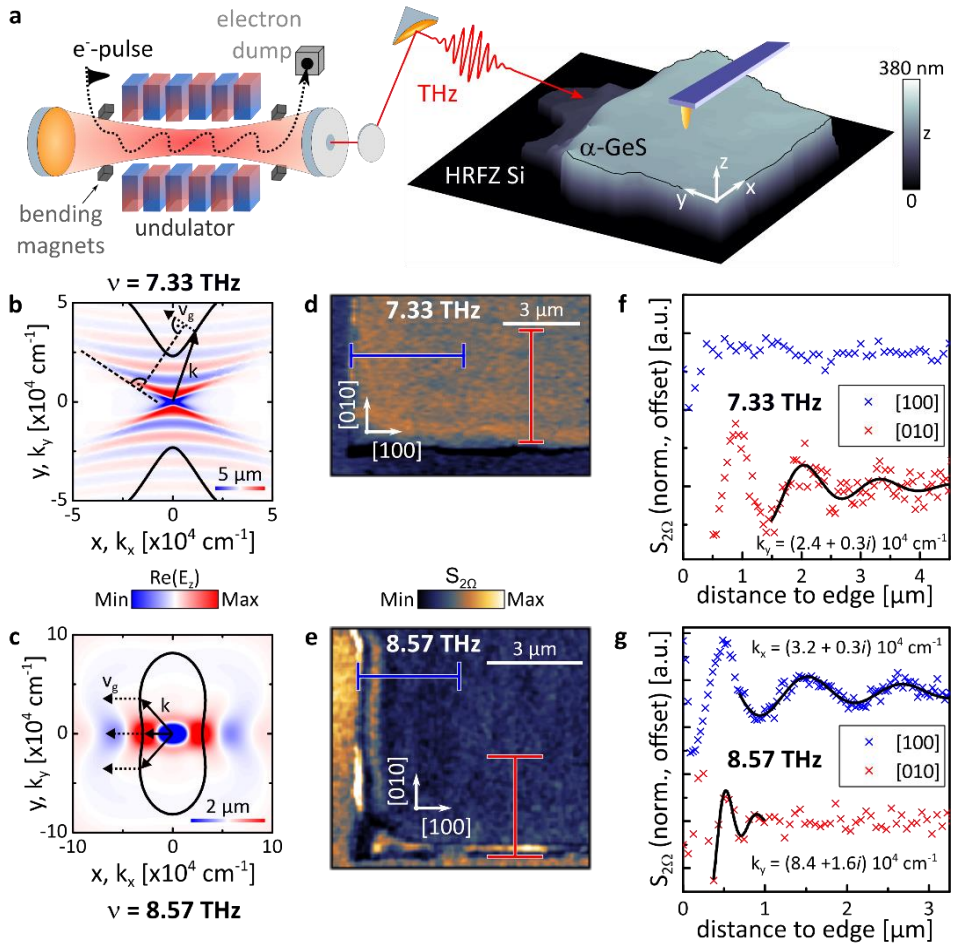

**Figure 2. Near-field imaging of THz PhPs in GeS. a)** Schematic of the experimental setup. The AFM tip excited by the FEL's THz radiation launches PhPs that propagate across a 224 nm-thick GeS slab. **b),c)** Simulated real-space field distributions Re($E_z$) at the GeS/air interface (false color plot, scale bar) overlaid with the isofrequency curves in momentum space (solid

black lines) at frequencies ν = 7.33 THz and ν = 8.57 THz, respectively. For selected k-vectors (solid arrows) the corresponding group velocities (dashed arrows) are depicted. In (b) the dashed line parallel to the group velocity limit is matching the opening angle in real-space. **d),e)** Optical near-field intensity $S_{2\Omega}$ images recorded in the bottom right corner of the slab at frequencies ν = 7.33 THz and ν = 8.57 THz, respectively. **f),g)** Averaged and high-pass-filtered near-field intensity $S_{2\Omega}$ profiles along the [100] (blue, curve offset for visibility) and [010] (red) crystal directions extracted from d) and e), respectively. The black solid lines are fits to the profile using Eq. (2) where applicable.

Finally, to determine the experimental polariton momentum along the [100] and [010] crystal directions from the two near-field images, we extract line profiles along the marked positions starting from the flake edges (blue and red, respectively in Figures 2d,e). The width of the marked profiles represents the area used for averaging, which is necessary to suppress the high noise level due to the high-frequency power fluctuations of the THz source. Additionally, the profiles are normalized and off-set for better visibility. The red profile along the [010]-direction in Figure 2f at ν = 7.33 THz clearly shows that PhPs decay exponentially with the distance to the GeS flake edge. This fitting (black curve in Figures 2d,e) was carried out using the common analytical description of an edge-reflected polaritonic wave[50]

$$f(x) = A \frac{1}{\sqrt{x}} \exp[-2 \operatorname{Im}(k)x] \sin[2 \operatorname{Re}(k)(x - x_0)] \qquad (2)$$

with arbitrary amplitude *A* and phase $x_0$. The result of the fit yields the complex momentum $k_{y,exp}^{7.33} = (2.4 + 0.3i) \times 10^4 \ cm^{-1}$ ($\lambda_{y,exp}^{7.33} = 2.6 \ \mu m$) that matches excellently to the values obtained from the simulation and analytical model. In the [100]-direction, no clear polaritonic features are visible, which is in line with the theoretical predictions, as well. The bright edge that appears in the corresponding NF image in Figure 2d is related to an edge effect due to far-field scattering off the GeS corner towards the substrate and hence does not appear in the profile for $x > 0$.

For the elliptical polariton at ν = 8.57 THz, both profiles in Figure 2g clearly show features of an exponentially decaying PhP electric field that can be fitted using Equation (2). We retrieve the momenta along the [100]- and [010]-direction to $k_{x,exp}^{8.57} = (3.2 + 0.3i) \times 10^4 \ cm^{-1}$ ($\lambda_{x,exp}^{8.57} = 2.0 \ \mu m$) and $k_{y,exp}^{8.57} = (8.4 + 1.6i) \times 10^4 \ cm^{-1}$ ($\lambda_{y,exp}^{8.57} = 0.75 \ \mu m$), respectively. Overall, we find a good agreement between the experimental data and the analytical model for $\mathbf{Re}(k_{x,y})$ along both directions. On the other hand, the experimental values for $\mathbf{Im}(k_{x,y})$ fall short of the theoretical values consistently by a factor of 3-4. Generally, the experimental error in Im(*k*) is considerably higher than Re(*k*) (here, due to the small amount of fitted oscillations and potential edge-launched contributions), but on average the experimental Im(*k*) is expected to be larger than the theoretical values, for example due to the decreased phonon life times (and coherence) in a real crystal (related to phonon scattering processes).

By recording near-field images at various illuminating frequencies and fitting the extracted $S_{2\Omega}$ profiles using **Equation (2)**, we obtain the PhP dispersion ν(*k*) directly from the experimental data in the frequency range ν = 6.0 – 8.7 THz (dots in **Figure 3**a). The left (right) panel relates to PhPs propagating along the [100] ([010])-direction, starting at the TO frequency ν$_{TO}$, where the permittivity becomes negative along the respective in-plane direction. Black curves show the corresponding PhP wavevectors Re[$k(\omega)$] calculated using Equation (1), in excellent agreement with the experiment (a minor adaptation to the GeS permittivity is carried out in RB$_x$, see Supporting Information). Moreover, we evaluated the reflectivity r$_p$(ν,k) of the layered air/GeS/Si-system via the transfer matrix formalism.[51] The Im[r$_p$(ν,k)] contains both the polariton dispersion and damping, with the positions of the maxima yielding the PhP dispersion and their width (FWHM) directly related to their damping. We find the Im[r$_p$(ν,k)] (false color plot in Figure 3a) matching excellently the experimental and analytical data, thus further corroborating our results.

Along the [100]-direction we find a phonon polariton branch emerging on the dispersion plot at ν$_{TO,[100]}$ = 7.74 THz: The momentum Re($k_x$) increases with frequency ν up to **Re($k_x$) = 0.4 × 10$^5$ cm$^{-1}$** (i.e., with a positive group velocity) with the relatively large width of the peak of Im[r$_p$(ν,k)] indicating considerable damping. Along the [010]-direction, in the frequency range ν = 6.06 – 7.9 THz, we comparably observe the polariton momentum Re($k_y$) to increase with frequency up to **Re($k_y$) = 0.55 × 10$^5$ cm$^{-1}$**, accompanied by a smaller damping as compared to the polariton branch along the [100]-direction. However, at higher frequencies ν > 7.9 THz, the behavior becomes much more intricate due to two consecutive back bending effects (anomalous dispersion) emerging in separate spectral areas. In general, back bending of a polariton's dispersion is well-known and can be due to several physical reasons: Firstly, surface plasmon or phonon polaritons show a similar effect at the spectral location where the area of negative permittivity ends.[52,53] In this case, the bending appears near the light line and the polariton branch emerging in the area with Re(ε) > 0 lacks interface confinement. Secondly, polaritons coupling to external excitations (such as phonons of the substrate[54] or nearby molecular resonances[10]) have been reported to induce a back bending or anti-crossing to the otherwise monotonous polariton dispersion. Finally, in the present case of a highly anisotropic material we find the anomalous dispersion to be induced by the PhPs coupling to intrinsic phonons. Such precondition is also met in α-MoO3, where a similar, considerably weaker effect was observed recently:[55] The weak [100]-phonon located at wavenumber ω$_{TO}$ = 998.7 cm$^{-1}$ couples to the in-plane elliptical PhP (that is caused by the negative permittivity in vdW-direction). As for the present GeS, here the PhPs coupling to the strong (weak) z-phonon with ν$_{TO,[001],1}$ = 7.1 THz (ν$_{TO,[001],2}$ = 8.4 THz) causes the back bending around ν = 8.5 THz (8 THz) (see Supporting Information). As seen in the reflectivity, the effect is accompanied by a substantial polariton damping that renders it challenging to observe experimentally with our current setup. Nevertheless, we measure the highest momenta **Re($k_y$) = 0.84 × 10$^5$ cm$^{-1}$** in between the regions that exhibit anomalous dispersion. In a nutshell, the highly anisotropic permittivity of GeS governed by overlapping degenerate optical phonon modes in a narrow spectral regime introduces an exotic in-plane PhP dispersion. The latter features several back-bending effects and three characteristic areas with different polariton modes that will be discussed below.

**Area A, 7.74 – 9.47 THz (ν$_{TO,[100]}$ - ν$_{LO,[010]}$):** In this spectral range, we find the permittivity Re(ε$_i$) to be negative along all three crystal directions and, hence, the polariton is constituted by the coupled interface modes[2] (air/GeS and GeS/Si). For a representative frequency ν = 8.35 THz, the in-plane IFC (Figure 3b), exhibits a propagating $l = 0$ mode (black curve). The shape relates to an in-plane elliptical PhPs in agreement with our experimental findings at ν = 8.57 THz (Figure 2e). In order to obtain the complete picture of the PhP field in three dimensions, we perform full-wave electromagnetic simulations of the layered system. Figure 3c displays a 3D representation of the vertical electric field component with the top face at a distance of 10 nm above the GeS slab, generated by the vertical electric dipole positioned $z = 265$ nm above the origin. The presented component Re($E_z$) is directly linked to the experiment as it provides a valid numerical description of the signals measured in s-SNOM.[56] As anticipated from the GeS permittivity, we find confined fields inside the GeS slab in both the x,z- and y,z-planes that are well confined to the GeS/Si interface and decay with $z$. Similarly, this interface PhP mode resides in the Si, featuring the same momentum ($k_\parallel$) as within the slab and decaying with decreasing $z$ (into the substrate). Overall, the polariton along the [010]-direction holds a considerably higher momentum Re($k_y$) > Re($k_x$) and a substantial damping Im($k_y$) > Im($k_x$), which is in agreement with our experimental findings at the frequency ν = 8.57 THz.

**Area B, 7.1 – 7.74 THz ($\nu_{TO,[001],1}$ - $\nu_{TO,[100]}$):** In this spectral range, GeS exhibits negative permittivity Re($\varepsilon_y$, $\varepsilon_z$) < 0 along both, the [010] and [001] crystal directions, whilst Re($\varepsilon_x$) > 0, which leads to the excitation of in-plane hyperbolic polaritons.[49] At ν = 7.5 THz the IFCs in Figure 3d support the hyperbolic character with the l = 0 mode (black curve) forming an open hyperbola with its major axis oriented along the [010]-direction. Of particular interest is the $l = 1$ mode (red curve) that likewise holds a hyperbola, which, however, displays its major axis aligned along the [100] direction. While different shapes of IFCs (and thus different directional properties) for the $l = 0$ and $l = 1$ modes have been theoretically predicted,[49] so far their experimental evidence is lacking. The simulation in Figure 3e supports the findings from the IFCs: The x,y-plane in air shows a hyperbolic PhP propagating along the [010] crystal direction. The field in the y,z-plane demonstrates the coupling of the interface modes. Within the x,z-plane, however, the field in air and Si is dominated by the non-polaritonic near field of the exciting dipole. Intriguingly, inside the GeS slab we find a high momentum PhP that decays along [100] direction rapidly with lateral distance x from the exciting dipole, while the propagating along this direction is forbidden for the zero order ($l = 0$) PhP slab mode, according to its IFC. We explain this observation by the propagation of the higher order ($l = 1$) PhP slab mode with its IFC shown in Figure 3d by the red curve.

**Area C, 6.1 – 7.1 THz ($\nu_{TO,[010]}$ - $\nu_{TO,[001],1}$):** At these frequencies only the [010]-crystal axis exhibits negative permittivity Re($\varepsilon_y$) < 0, resulting in a type II hyperbolic optical response with hyperbolic PhP slab modes.[2] At ν = 6.8 THz the IFCs (Figure 3f) show the peculiar in-plane hyperbolic shape with the hyperbola's major axis aligned along [100]-direction for both propagating modes ($l = 0, 1$). Compared to ν = 7.5 THz (Area B), we find a smaller Re($k_x$) as anticipated from the dispersion in Figure 3a. The simulated fields in Figure 3g within the x,y-plane similar to that in Area B depict hyperbolic PhPs with low momentum $k_y$ propagating in [010]-direction in agreement with the the IFC of the $l = 0$ PhP mode (black curve in Figure 3f). The fields in the x,z-plane solely show the non-PhP near fields of the exciting dipole along [100]-direction as PhP propagation in this direction is forbidden. In contrast, the field distribution within the y,z-plane depicts a polariton with small damping propagating along the [010]-direction. The remarkably high momentum $l = 1$ mode predicted by Equation (1) cannot be easily observed.

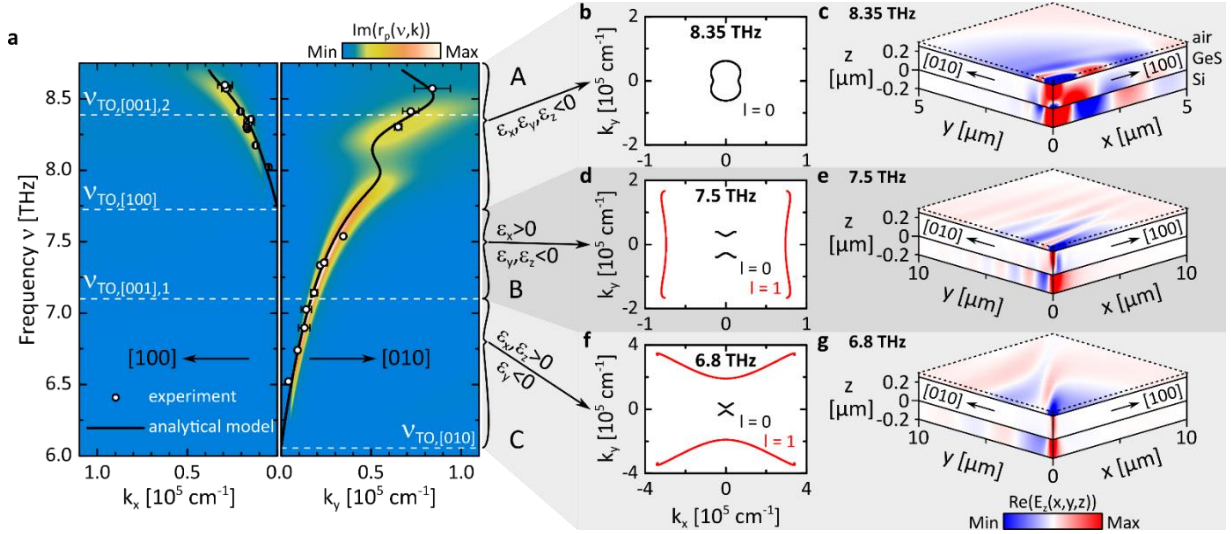

**Figure 3. Dispersion and field distributions of PhPs in GeS. a)** Dispersion ν(k) along the [100] (left side) and [010] (right side) crystal directions for a 224 nm-thick GeS slab. The dots represent the experimental data extracted from near-field profiles and the black curve corresponds to eq. (1) with $l = 0$ and φ = 0, π/2. The false color plot presents the imaginary part of the reflection coefficient $r_p$(ν,k) calculated via the transfer-matrix formalism.[51] **b),d),f)** IFCs calculated for three characteristic frequencies ν = 8.35, 7.5, and 6.8 THz within the different reststrahlen bands, respectively. The black (red) curves relate to the propagating $l = 0$ ($l = 1$) modes. **c),e),g)** 3D representation of the simulated field distributions Re($E_z$) at the same three frequencies for a 224 nm-thick GeS slab between air and an Si substrate. The top face relates to the x,y-plane as observed in the experiment and the right (left) face corresponds to the y,z- (x,z-) face.

The primary difference between the PhPs in areas B and C is presented by the rotation of the hyperbolic $l = 1$ mode, as the $l = 0$ mode proves to be unaffected by the change of sign in Re($ε_z$). A more detailed analysis of the transition at the frequency $ν_{TO,[001],1}$ can be found in the Supporting Information.

Ultimately, the aim of this work is to thoroughly characterize the PhPs in GeS to pave the way towards application in science and technology. To this end, we determine the key GeS polaritonic properties (figure of merit, life time τ, and light confinement β) from our experiment as well as the analytical model and compare them to recent, established PhP-hosting materials. The quality factor Q = (Re(k))/(Im(k)) presents a practical figure of merit (FOM) that (in real space) relates the polariton's wavelength to its decay length.[6] For GeS, in **Figure 4**a we find FOMs of up to Q = 10 in $RB_y$ and Q = 3 in $RB_x$ along the [010] and [100] directions, respectively. The black curves are obtained directly from Equation (1) and describe well the experimental data. The data points distinctly falling below the model can be attributed to an increased bandwidth of the exciting FEL pulse that leads to an artificially increased polariton damping: in fact, an incident pulse with significant bandwidth according to the dispersion in Figure 3a excites a broad range of polaritons with various k-vectors. The super-position of such waves detected in the far-field exhibits a shortened propagation length due to destructive interference (more detailed explanation given in the Supporting Information). Note that the quality factor drops at the frequencies of the in-plane LO and TO as well as in the regions of the back-bending in the dispersion. Overall, the quality factors resemble those reported for α-MoO$_3$ (Q ≈ 7 - 12) and are 2-times smaller than in naturally abundant hBN (Q ≈ 20)[20] and about 3-times higher than in α-V$_2$O$_5$ (Q = 2.5).[5]

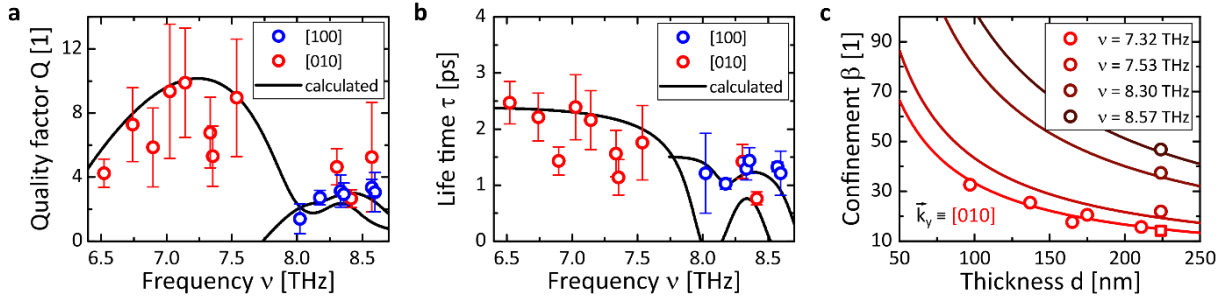

**Figure 4: Characteristic properties of PhPs in GeS. a)** Experimental quality factors as a function of frequency for polaritons propagating along the [100] (blue) and [010] (red) crystal directions. The two black curves are the analytical estimates obtained from eq. (1) for the in-plane crystal axes (φ = 0, π/2 and $l$ = 0). **b)** Polariton life time as a function of frequency along the in-plane crystal axes. Again, the black curves are obtained from Equation (1). **c)** Thickness-dependence of the polariton field confinement at four different frequencies within RB$_y$. The experimental data taken at ν = 7.32 THz for a set of different GeS flakes follow the well-known ~1/$d$ behavior in eq. (1) (solid lines). The squared data point was taken at ν = 7.33 THz on the $d$ = 224 nm-thick flake.

The GeS PhPs life time ($\tau = [v_g \, \text{Im}(k)]^{-1}$ with the group velocity $v_g = 2\pi c \, d\omega / dk$, ω denoting the wavenumber) lies in the picosecond range (Figure 4b) as anticipated for PhPs.[57] Within RB$_y$, life times of up to $\tau_{[010]}$ = 2.3 ps can be found, while the life times in RB$_x$ are considerably smaller with $\tau_{[100]}$ < 1.4 ps. The life times are thus comparable with those reported for PhPs in hBN (< 2 ps),[20] but shorter than in case of PhPs in α-MoO$_3$ (2 - 8 ps)[4] and α-V$_2$O$_5$ (3 - 6 ps).[5] Note that in the same way as for the FOM, a higher excitation bandwidth can artificially decrease the experimental life time (see Supporting Information). Moreover, the large errors in the determination of life time as well as Q factor are related to the poor signal to noise ratio within the experiment, which is appositely backed up by our simulations: For GeS, we find a smaller overall polaritonic field Re($E_z$) per exciting field strength as compared to α-MoO$_3$ and hBN, for example. Moreover, it is worth mentioning that following the common definition of the life time (propagation length $L = \text{Im}(k)^{-1}$ divided by the group velocity $v_g$) can erroneously lead to negative values (in the regions of anomalous dispersion, where $d\omega / dk < 0$) as for instance in the dispersion curve along [010]-direction in Figure 3a. In the anomalous dispersion region, one has to use a different (more general) determination of the life time, based on the eigenmode analysis in the space of complex frequency and real wavevector.[10]

Finally, we calculate the thickness-dependent light confinement $\beta = k / k_0 = \lambda_0 / \lambda$ (that is, the ratio of the polariton in-plane momentum $k$ in respect to the incident, free-space light $k_0$) at different frequencies in RB$_y$. As presented in Figure 4c, the experimental values of $\beta$ follow very well the ~1/$d$-dependence anticipated from Equation (1) (solid curves). The squared data point is taken from the dispersion of the 224 nm-thick flake in Figure 3a at a slightly shifted frequency of ν = 7.33 THz. We find in our experiment the highest field confinement of $\beta$ = 47 in the 224 nm-thick flake at ν = 8.57 THz, whereas considerably larger values are expected for thinner GeS flakes.

Eventually, we want to address the apparent PhP canalization indicated by the numerical simulation and analytical model presented in Figure 2e. Similar effects have been studied mainly theoretically in the context of plasmonic and phononic metamaterials[19,58,59] and, to our knowledge, have not been experimentally demonstrated in a single layer of a natural material yet. In GeS, the canalization is caused by the anisotropic permittivity that leads to the propeller-shaped, almost flat IFCs together with a large damping along the [010]-direction, resulting in propagating polaritons with nearly parallel group velocities. Hence, the wave fronts lose their hyperbolic character as observed in the Re($E_z$) image. Direct experimental observation of this (nearly) diffraction-less propagation will require an antenna for efficiently launching these polaritons in order to probe this electric field distribution emitted by an effective point source. Hence, a dedicated, frequency-dependent study of this canalization effect will be presented in a future work.

In conclusion, we extensively explored the properties of THz phonon polaritons in the highly anisotropic semiconducting vdW material α-GeS, finding different confined polaritonic modes between frequencies of 6 and 9 THz. Of particular interest are the anomalous dispersion and the anticipated natural canalization effect that both originate from the interplay of the highly directional RBs. For these reasons, GeS promises to become a feasible, versatile platform for THz light confinement in the future. Moreover, we believe the work presented here will inspire novel research on THz PhPs: While on one hand, the material family presents a toolbox for THz PhP engineering (for example via stacking and twisting), on the other hand, GeS holds the promise of tuning the PhPs via electrical gating (i.e. PhP-electron interaction). Direct control of the charge carrier concentration moreover enables the study of plasmon-phonon coupling with the goal to actively control the anisotropic polaritons. Lastly, the large thermoelectric effect motivates investigation of the thermoelectric properties of the PhPs that could potentially be probed via photocurrent-nanoscopy[26]. To that end, the scarcity of tunable THz sources currently presents the only limitation.

**Experimental section**

*Scattering-type scanning near-field optical microscopy (s-SNOM):*

The near-field images were recorded applying a (modified) commercial near-field microscope (Neaspec GmBH, Germany) conjoined with the free-electron laser located at Helmholtz-Zentrum Dresden-Rossendorf (HZDR), Germany. By illuminating the oscillating ($\Omega$ = 250 kHz), metallized s-SNOM-tip in the vicinity of the sample surface, the excited tip acts as an antenna providing a strong, localized electric field at its apex. The confined field interacts with the sample volume and hence, its local optical response becomes imprinted in the back-scattered signal $S$. In order to separate the sample's near-field optical response from the dominant far-field background, the non-linear distance-dependence of the NF contribution (as compared to the linear dependence of the far-field) is exploited:[60] Through employing a lock-in amplifier we obtain individual components of the scattered signal at multiples of the cantilever oscillation frequency $S_{n\Omega}$ ($n$ = 1, 2, 3…) and find effective background suppression in the components with $n \geq 2$.[61] Throughout this work, a self-homodyne detection scheme was applied and the back-scattered optical signal was demodulated at $n$ = 2. The optical signal at the frequencies $\nu$ = 6 - 9 THz was recorded using a gallium-doped germanium photoconductive detector by QMC Instruments Ltd., UK.

*Free-electron laser:*

Light sources in the THz spectral regime that are suitable for s-SNOM application currently present a major limitation to near-field optical investigation of collective excitations in condensed matter physics. Established table-top solutions such as gas-lasers or quantum cascade lasers are either restricted by the small range of accessible frequencies or the lack of sufficient spectral power density.[62] In contrast, light emission of relativistic electrons can be exploited in large-scale facilities (namely synchrotrons or free electron lasers) to provide either broadband (in the first case) or continuously tuneable, narrowband THz radiation (in the latter case). While synchrotron infrared nanospectroscopy currently is operational at frequencies down to > 9.6 THz,[25,63] FELs in particular have been successfully applied in s-SNOM from 1.3 – 30 THz.[24,64]

In this work, we apply the free-electron laser FELBE at the ELBE Center for High Power Radiation Sources at HZDR, Germany, capable of generating coherent THz radiation over the spectral range of 1.2 – 60 THz with a repetition rate of 13 MHz. Particularly the U100 FEL oscillator provides the required brightness to launch and detect PhPs in the 6 – 9 THz spectral regime. The spectral bandwidth of individual pulses was minimized by slightly detuning the cavity, resulting in values of about 0.5 – 0.9%$_{FWHM}$ and transform-limited pulse durations of > 5 ps. The implied pulse spectral diagnostic was performed applying a Czerny-Turner type scanning grating spectrometer (Princeton Instruments SP-300i).

*Full-wave numerical simulations*:

The structures were modelled as biaxial GeS slabs on top of high-resistivity float-zone Si substrates. In s-SNOM experiments the tip acts as an optical antenna that converts the incident light into a strongly confined near field below the tip apex, providing the necessary momentum to excite PhPs. However, owing to the complex near-field interaction between the tip and the sample, numerical quantitative studies of s-SNOM experiments meet substantial difficulties in simulating near-field images.[65] To overcome these difficulties, we approximate the tip by a dipole source (with a constant dipole moment),[56] in contrast to the usual dipole model, in which the effective dipole moment is given by the product of the exciting electric field and the polarizability of a sphere.[66] We assume that the polarizability of the dipole is weakly affected by the PhPs excited in the GeS slab, and their back-action onto the tip can be thus neglected. Therefore, we place a vertically oriented point electric dipole sources on top of the GeS slab and calculate the amplitude of the near field, $|E_z|$, above the GeS/Si structure, where PhPs propagate. Our simulated images (using Comsol Multiphysics) are in good agreement with our experimental results (see Figure 3), which lets us conclude that the calculated field between the dipole and the GeS flake, $E_z$, provides a valid numerical description of the signals measured by s-SNOM.

**Acknowledgements**

T.N., M.O., L.W., S.C.K., L.M.E., and T.V.A.G.O. acknowledge the financial support by the Bundesministerium für Bildung und Forschung (BMBF, Federal Ministry of Education and Research, Germany, Project Grant N°s 05K16ODA, 05K16ODC, 05K19ODA, and 05K19ODB) and by the Deutsche Forschungsgemeinschaft (DFG, German Research Foundation) under Germany's Excellence Strategy through Würzburg-Dresden Cluster of Excellence on Complexity and Topology in Quantum Matter - ct.qmat (EXC 2147, project-id 390858490). F.H. gratefully acknowledges financial support by the Deutsche Forschungsgemeinschaft (DFG) through the project CRC1415 (ID: 417590517). A.Y.N. acknowledges the Spanish Ministry of Science and Innovation (grants MAT201788358-C3-3-R and PID2020-115221GB-C42) and the Basque Department of Education (grant PIBA-2020-1-0014). G.Á.-P. acknowledges support through the Severo Ochoa Program from the Government of the Principality of Asturias (grant numbers PA-20-PF-BP19-053). P.A.-G. acknowledges support from the European Research Council under starting grant no. 715496, 2DNANOPTICA and the Spanish Ministry of Science and Innovation (State Plan for Scientific and Technical Research and Innovation grant number PID2019-111156GB-I00).